\documentclass[nohyperref]{article}
\usepackage{amsmath}
\usepackage{bbm}
\usepackage{natbib}
\usepackage{tikz}
\usepackage[utf8]{inputenc}

\usepackage{microtype}
\usepackage{graphicx}
\usepackage{subfigure}
\usepackage{booktabs} 
\usepackage{ulem} 

\usepackage{hyperref}

\usepackage[accepted]{icml2022}

\newtheorem{theorem}{Theorem}

\title{\textsc{Are Synthetic Control Weights Balancing Score?}}
\author{\textsc{Harsh Parikh\footnote{harsh.parikh@duke.edu, \textsc{Duke University}} }}
\date{}

\begin{document}

\maketitle

\begin{abstract}
     In this short note, 
     I outline conditions under which conditioning on Synthetic Control (SC) weights emulates a randomized control trial where the treatment status is independent of potential outcomes. Specifically, I demonstrate that if there exist SC weights such that (i) the treatment effects are exactly identified and (ii) these weights are uniformly and cumulatively bounded, then \textit{SC weights are balancing scores}.
\end{abstract}

\section*{Introduction}
Synthetic control (SC) and balancing score methods are critical tools for causal inference \cite{abadie2010synthetic,abadie2015comparative, rosenbaum1983central,hansen2008prognostic}. The theoretical properties of these approaches are well-studied in the literature \cite{abadie2015comparative, ferman2016revisiting, ferman2017placebo, shi2021theory, rosenbaum1983central, hansen2008prognostic}. However, the connections between them are not well understood. This short note bridges this gap by proving that SC weights are balancing scores under a sufficient set of assumptions. It is of general interest because conditioning on balancing scores guarantees ignorability (i.e., that potential outcomes and treatment assignment are mutually independent) -- ensuring consistent estimation of treatment effects \cite{johnson2013compared}.

\subsection*{Setup}\label{sec:expo}
Consider a finite population of $n$ units, denoted by $\mathcal{S}_n$. 
For each unit $i \in \mathcal{S}_n$, we observe the following three quantities: (i) $Z_i$, the binary treatment indicator, (ii) $\{X_{i,t}\}_{t=0}^{t_0}$, the time-series of pre-treatment outcomes, and (iii) $\{Y_{i,t}\}_{t=t_0+1}^{T}$, the time-series of post-treatment outcomes. Our setup assumes that exactly one of the units is treated. Without loss of generality, let $Z_1 = 1$ and $\forall j>1, Z_j=0 $. We assume that these timeseries are generated using the following factor model:
\begin{eqnarray}
    \text{for} \; t \leq t_0:&& \;\; X_{i,t} = \delta_t + \lambda_t \mu_i + \epsilon_{i,t} \\
    \text{for} \; t > t_0:&& \;\; Y_{i,t} = \delta_t + \lambda_t \mu_i + Z_i \alpha_{i,t} + \epsilon_{i,t} 
\end{eqnarray}
where, (i) $\delta_t$ is an unobserved common time-trend across units, (ii) $\mu_i$ is an unobserved unit specific factor, (iii) $\lambda_t$ is the time specific factor loading, (iv) $\epsilon_{i,t} \sim \mathcal{N}(0,\sigma_{i,t}^2)$ is the noise at time $t$ and (v) $\alpha_{i,t}$ is the treatment effect for unit $i$ at time $t$. We are interested in identifying $\{\alpha_{1,t}\}_{t=t_0+1}^T$. This setup is similar to the one discussed in \cite{ferman2017placebo}. 

\begin{figure}
    \centering
    \includegraphics[width=0.48\textwidth]{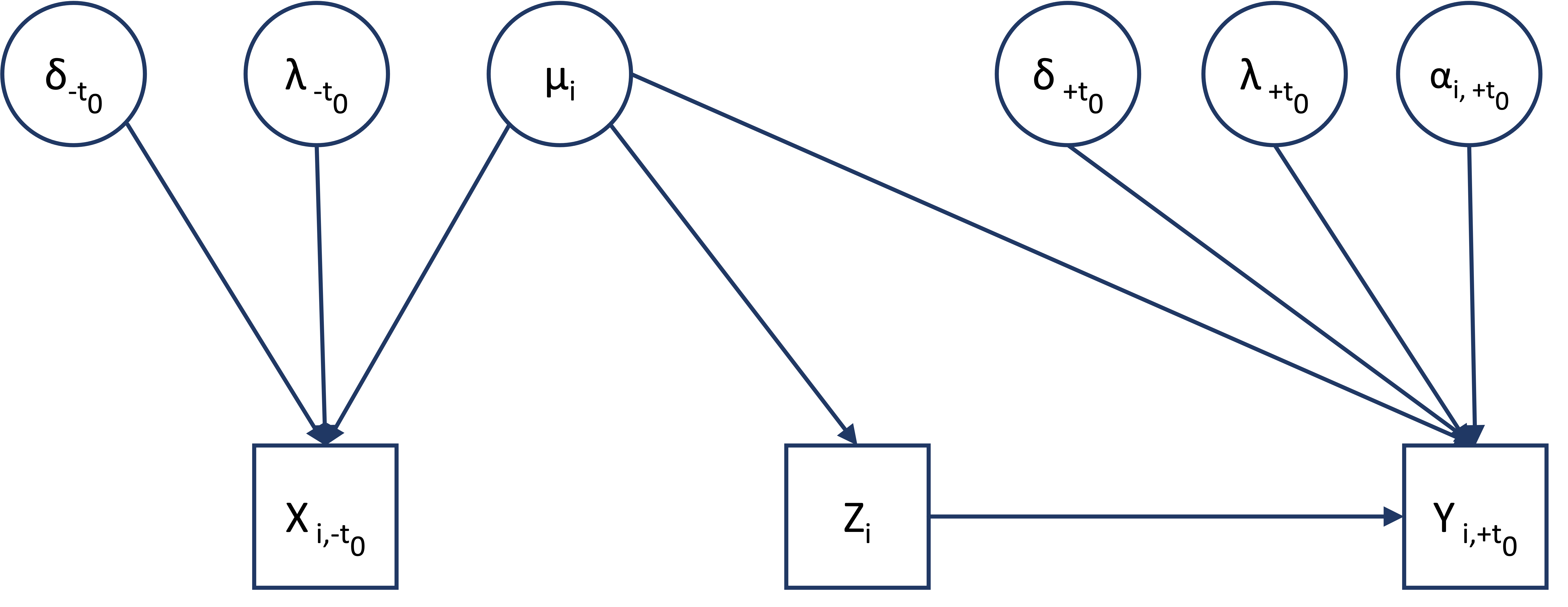}
    \caption{A simplified causal directed acyclic graph for the discussed exposition. Here, with the slight abuse of notation, $-t_0$ refers to the time before intervention and $+t_0$ refers to the time after intervention.}
    \label{fig:dag}
\end{figure}

We will use bold letters to denote the collection of observations across units: $\boldsymbol{Z} = \{Z_i\}_{i=1}^{n}$, $\boldsymbol{X} = \{\{X_{i,t}\}_{t=0}^{t_0}\}_{i=1}^{n}$ and $\boldsymbol{Y} = \{\{Y_{i,t}\}_{t=t_0+1}^{T}\}_{i=1}^{n}$. Consider, the scenario where the probability of observing treatment assignments is a function of the unit specific factors $\boldsymbol{\mu} = \{\mu_i\}_{i=1}^{n}$ i.e.
\begin{equation}
    \boldsymbol{Z} \sim f(\boldsymbol{\mu})   
\end{equation}
for some (unknown) distribution $f$ such that $\mathbf{1}^T\boldsymbol{Z} = 1$. Figure~\ref{fig:dag} graphically demonstrates the causal dependencies described in equations 1, 2 and 3.

\subsection*{Synthetic Control: Assumptions and Identification}
Consider the following assumption:\\
\textit{A.1. (\textbf{feasibility}) $\exists \boldsymbol{\beta}$ s.t. $\mathbbm{E}\left(X_{1,t} - \sum_{i=2}^n \beta_i X_{i,t}\right) = 0$.}

\textbf{Identification}. Assumption A.1. implies that $\mu_1 = \sum_{i=2}^n \beta_i \mu_i$. If we choose $\beta_i$'s such that the condition in A.1. are satisfied, then 
\begin{equation*}
    \mathbbm{E}\left(Y_{1,t} - \sum_{i=2}^n \beta_i Y_{i,t}\right)  = \lambda(\mu_1 -  \sum_{i=2}^n \beta_i \mu_i) + \alpha_{1,t} = \alpha_{1,t}.
\end{equation*}
Thus, identifying $\boldsymbol{\beta}$ such that $\mathbbm{E}\left(X_{1,t} - \sum_{i=2}^n \beta_i X_{i,t}\right) = 0$ leads to the identification of the treatment effect $\alpha_{1,t}$.


We, further, assume that there exists \textit{feasible} weights $\boldsymbol{\beta} = \{\beta_2,...,\beta_n\}$ such that:\\
    \textit{A.2. (\textbf{uniformly bounded}) $\forall i$,  $0\leq\beta_i\leq1$,}\\
    \textit{A.3. (\textbf{cumulatively bounded}) $\sum_{i=2}^n \beta_i = 1$ i.e. $\mathbf{1}^T\boldsymbol{\beta}=1$.}


We refer to these as \textit{oracle weights} because it may not be feasible to exactly identify them using the observed finite population, especially, when the noise is heteroskedastic\footnote{To estimate, these weights, in practice, one fits a model that uses the outcomes of the control units to predict the contemporary outcome of the treated unit in the pretreatment period: $\texttt{regress y=} X_{i,t}, \texttt{x=}\{X_{2,t}\dots X_{n,t}\}$. Typically, the regularization ensures that the weights are bounded between 0 and 1 and the sum of the weights is one.} \cite{ferman2017placebo}. For the rest of the argument, we will assume that we have the knowledge of these oracle weights. However, one can always estimate approximate oracle weights by fitting a regularized linear regression on the pre-treatment outcomes \cite{abadie2010synthetic, abadie2015comparative}.



\section*{SC Weights are Balancing Score}
For our setup, if there exists a function $b$ such that $\{\boldsymbol{Y}_t(\boldsymbol{z})\}_t \perp \boldsymbol{Z} | b(\mathbf{X})$ then $b$ is a \textit{balancing score}. However, consider the causal graph in Figure~\ref{fig:dag}: it is not obvious that $b(\mathbf{X})=\mathbf{X}$ is a balancing score. 

We know that $\{\boldsymbol{Y}_t(\boldsymbol{z})\}_t \perp \boldsymbol{Z} | \boldsymbol{\mu}$; thus, $\boldsymbol{\mu}$ is a balancing score. However, one must note that $\boldsymbol{\mu}$ is unobserved and hence it is not a very useful balancing score. In this discussion, we show that the (oracle) SC weights are also balancing scores i.e. $\{\boldsymbol{Y}_t(\boldsymbol{z})\}_t \perp \boldsymbol{Z} | \boldsymbol{\beta}$. 

\begin{figure}
    \centering
    \includegraphics[width=0.23\textwidth]{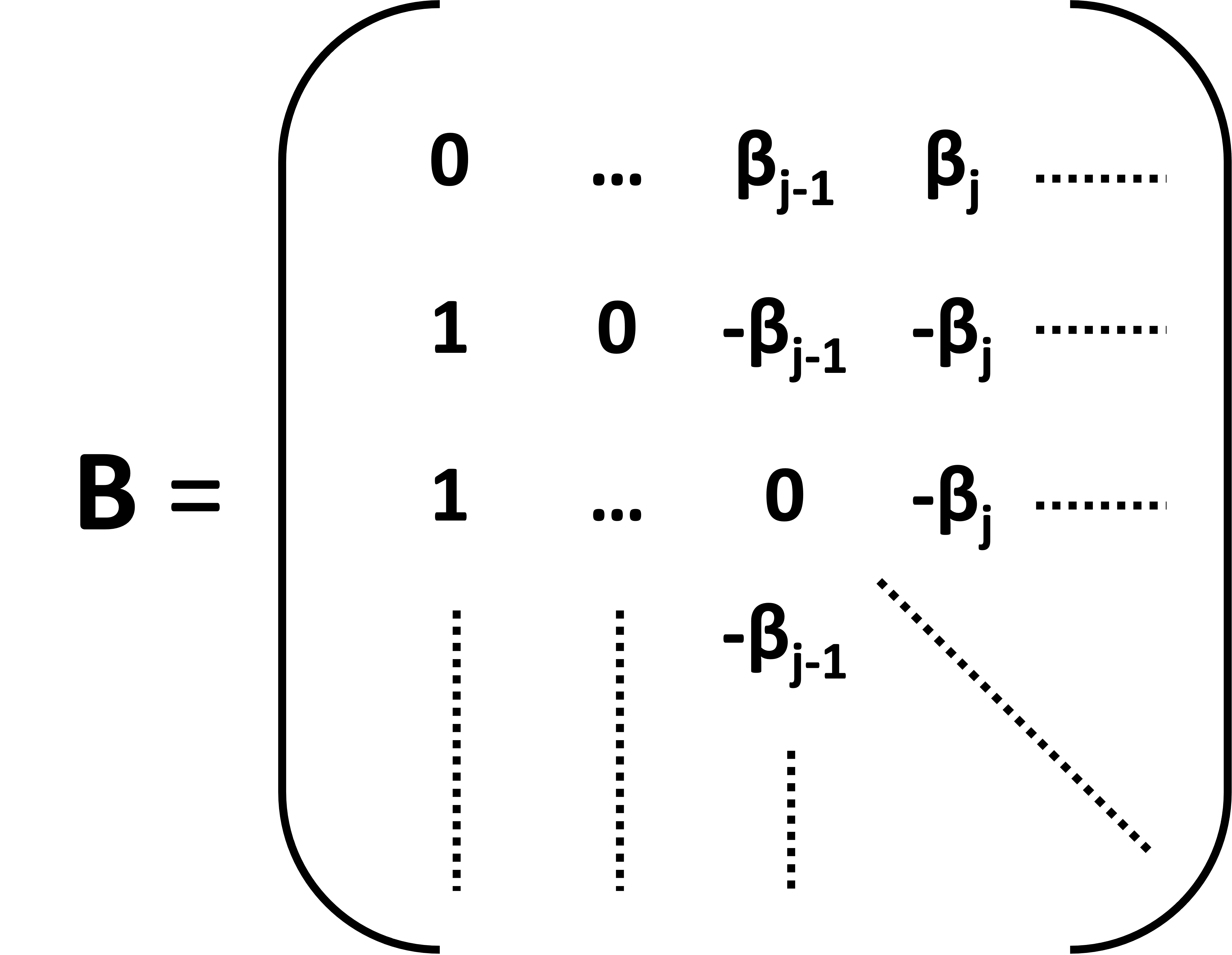}
    \caption{Construction of $\mathbf{B}$ matrix such that $\mathbf{B}_{i,i}=0$, $\mathbf{B}_{i,1}=1$ (for $i>1$), $\mathbf{B}_{1,j}=\beta_j$ (for $j>1$) and $\mathbf{B}_{i,j}=-\beta_j$ (for $j>1$ and $i>1$). }
    \label{fig:B}
\end{figure}

\begin{theorem}
Given assumptions A.1 - A.3,
$$\{\boldsymbol{Y}_t(\boldsymbol{z})\}_t \perp \boldsymbol{Z} ~|~ \boldsymbol{\beta}$$.
\end{theorem}
\vspace{-6mm}\textbf{Proof}. Consider an $n\times n$ matrix $\mathbf{B}$ such that: (a) $\mathbf{B}_{i,i}=0$, (b) $\mathbf{B}_{i,1}=1$ (for $i>1$), (c) $\mathbf{B}_{1,j}=\beta_j$ (for $j>1$), and (d) $\mathbf{B}_{i,j}=-\beta_j$ (for $j>1$ and $i>1$) (see Figure~\ref{fig:B}). Now, \textit{we will show}, $\{\boldsymbol{Y}_t(\boldsymbol{z})\}_t \perp \boldsymbol{Z} | \mathbf{B}$. We know that $\mathbbm{E}(\boldsymbol{X}_t) = \delta_t + \lambda_t \boldsymbol{\mu}$, $\mathbbm{E}(X_{1,t}) = \sum_{i=2}^n \beta_i \mathbbm{E}(X_{i,t})$ and $\sum_{i=2}^n \beta_i = 1$. These results imply that $\mu_1 = \sum_{i=2}^n \beta_i \mu_i$. We now observe that $\boldsymbol{\mu} = \mathbf{B}\boldsymbol{\mu}$, i.e. $\boldsymbol{\mu}$ is one of the eigenvectors of the matrix $\mathbf{B}$ with corresponding eigenvalue equal to 1. Hence, $\{\boldsymbol{Y}_t(\boldsymbol{z})\}_t \perp \boldsymbol{Z} | \mathbf{B}$ and $\mathbf{B} = g(\boldsymbol{\beta})$ where $g$ is the discussed matrix construction. \textit{QED}.

\textit{Note} that as $\boldsymbol{\beta}$ is a deterministic function of $\mathbf{X}$, $\mathbf{X}$ is also a balancing score i.e.  $\{\boldsymbol{Y}_t(\boldsymbol{z})\}_t \perp \boldsymbol{Z} | \mathbf{X}$. 



\section*{Conclusion and Discussion} \label{sec:conclude}
 In this short note, we discuss the set of sufficient conditions under which SC weights are balancing scores. Learning that SC weights are balancing scores allows for theoretical study of SC methods from a new lens. This understanding may allow for answering questions about the validity of SC inferences. Further, a researcher can borrow methods from one domain to bolster the estimation approach in a separate domain. For instance, one can think of a doubly robust method for time-series data where (generative) time-series models are augmented with SC weights for efficient and robust estimation. More work is needed to understand if this result holds in different scenarios where synthetic controls have been used for estimation. Further, it will be important to understand the validity of estimation when the SC weights are not balancing scores.
 \subsection*{Acknowledgements}
 I want to thank Profs. Alexander Volfovsky, Eric Tchetgen Tchetgen and Cynthia Rudin for their comments to improve this note. 
\bibliographystyle{icml2022}
\bibliography{biblio}
\end{document}